# Synthetic Minority Over-sampling TEchnique (SMOTE) for Predicting Software Build Outcomes

Russel Pears & Jacqui Finlay
School of Computing & Mathematical Sciences
Auckland University of Technology
Auckland, New Zealand
russel.pears@aut.ac.nz

Andy M. Connor
Colab
Auckland University of Technology
Auckland, New Zealand
andrew.connor@aut.ac.nz

*Abstract*— **In this research we use a data stream approach to mining data and construct Decision Tree models that predict software build outcomes in terms of software metrics that are derived from source code used in the software construction process. The rationale for using the data stream approach was to track the evolution of the prediction model over time as builds are incrementally constructed from previous versions either to remedy errors or to enhance functionality. As the volume of data available for mining from the software repository that we used was limited, we synthesized new data instances through the application of the SMOTE oversampling algorithm. The results indicate that a small number of the available metrics have significance for prediction software build outcomes. It is observed that classification accuracy steadily improves after approximately 900 instances of builds have been fed to the classifier. At the end of the data streaming process classification accuracies of 80% were achieved, though some bias arises due to the distribution of data across the two classes over time.**

## I. INTRODUCTION

The Mining Software Repositories (MSR) field analyses the rich data available in software repositories to uncover interesting and actionable information about software systems and projects. This enables researchers to reveal interesting patterns and information about the development of software systems. MSR has been a very active research area since 2004 [4]. Until the emergence of MSR as a research endeavor, the data from software repositories were mostly used as historical records for supporting development activities. Analysis of MSR research has shown that the approach of extracting knowledge from a repository has the potential to be a valuable method for analyzing the software development process for many domains [5]. However there are a number of data related challenges, one of which is how to deal with repositories that contain insufficient or imbalanced data because the project is still immature or because the development environment is such that data is discarded over time.

Our previous work [21] developed an approach for applying data stream mining techniques to overcome the challenge of discarded data utilizing the Jazz repository [1]. The Jazz repository stores data, including source code, related to each software build attempt and retains the build outcome, categorized as success or failure. As the volume of data associated with each build is large only a limited number of build instances are actually stored in the repository on a first-in, first-out basis. Traditional data mining methods are tailored to static data environments where the data is retained. The first major challenge in mining software repositories is dealing with dynamic data that arrives on a continuous basis. Our previous work [21] addressed this challenge by modeling the development process as a data stream to deal with software project data that is produced continuously and accumulated over a period of time before being discarded. The data stream mining approach was shown to be effective at maintaining knowledge related to the development project even after the data is discarded.

This paper attempts to extend our work to address the second challenge, namely to deal with the limited volume of data and improve the accuracy of the prediction event. In order to boost the training power of the limited quantity of available data the SMOTE [2] oversampling algorithm was applied to synthesize new data instances from the available instances prior to inducing a decision tree model implemented via the Hoeffding [3] tree method. For the simulation to be realistic the naturally occurring distribution of successful and failed build instances occurring in the original population was maintained in the oversampling process.

The dynamic nature of software and the resulting changes in software development strategies over time causes changes in the patterns that govern software project outcomes. This phenomenon has been recognized in many other domains and is referred to as concept drift. Changes in a data stream can evolve slowly or quickly and rates of change can be queried within stream-based tools. This paper describes an attempt to improve build outcome prediction accuracies for the Jazz project by synthetically creating data to boost the training power of the data stream mining approach while taking into account concept drift that occurs as part of the stream.

## II. BACKGROUND AND RELATED WORK

This research draws from multiple areas to inform the direction of inquiry, in particular it is placed in the context of other research related to Mining Software Repositories research, specifically in the context of the Jazz repository. In addition, it uses experience gained applying data stream mining and synthetic data generation in other domains to improve the prediction models developed for the Jazz project.

*A. Mining the Jazz Repository*

The Jazz development environment has been recognized as offering new opportunities in terms of MSR research because it integrates the software source code archive and bug database by linking bug reports and source code changes with each other [6]. Whilst this provides much potential in gaining valuable insights into the development process of software projects, such potential is yet to be fully realized. To date, much of the work focused on the Jazz repository is related to predicting build success, either through social network analysis [7] or source code metrics [21, 22]. As is common with much MSR research, the goal of working with the Jazz repository is in line with a key direction identified in the field [23], which is the transformation of software repositories from static record-keeping ones into active repositories in order to guide decision processes in modern software projects.

*B. Data Stream Mining*

The mining of data streams has arisen as a necessity due to advances in hardware/software that have enabled the capture of different measurements of data in a wide range of fields [24]. Data streams are typically generated continuously and have very high fluctuating data rates. The storage, querying and mining of such data sets are computationally challenging tasks [24]. Research problems and challenges that have been arisen in mining data streams can be solved using well-established statistical and computational approaches that can be categorized as either data-based or task-based ones. In data-based solutions, only a subset of the whole dataset is examined or the data is transformed to an approximate smaller size representation. Task-based solutions involve applying techniques from computational theory to achieve time and space efficient solutions. Data-based solutions include Sampling, Load Shredding, Sketching and Aggregation. Task-based solutions include Approximation Algorithms and Sliding Window approaches, all of which have received considerable attention by researchers [28]. The discarding of data from the Jazz environment and the relatively low data rate lends itself to a Sliding Window solution.

In addition, various data mining approaches can be applied to mining data streams, including clustering, frequency counting and classification. The nature of the data, which includes a classifiable attribute in terms of build outcome, lends itself to a classification method. In this work, we have applied the Hoeffding tree incremental learner in conjunction with the Adaptive Sliding Window (ADWIN) concept drift detector. ADWIN is a parameter-free adaptive sliding window drift detector that compares all adjacent sub-windows in given data window in order to detect a concept drift point [29]. This method is recognized to produce high true positive and low false positives rates while, having low detection delay times in comparison to other drift detectors proposed in the data mining literature [29].

*C. Synthetic Data Generation*

Many of the challenges associated with data stream mining are related to dealing with high volumes of data in relatively short timescales. It therefore seems counter-intuitive to deploy synthetic data generation techniques in conjunction with a data stream mining approach. However, the discarding of data from Jazz does not encourage the use of static classification approaches in practice, even though such approaches can be deployed on any given snapshot of the repository [22]. Deploying a data stream method in conjunction with synthetic data generation allows a consistent approach to be used in practice for new projects. Synthetic data can be generated from the limited quantity of actual data that is available in the early stages of development, and a gradual phasing out of such synthetic data can be carried out when larger volumes of real data become available. Such consistency is important if data mining approaches are to become useful to software practitioners.

Synthetic data generation has been a research area for some time, with the literature containing many examples of random or pseudo-random data generation [25]. However, the goal of our research is such that synthetic data must be representative of the real data and therefore a more refined generation approach is required. Such approaches include DataBoost-IM [26], ADASYN [27] and SMOTE [2] to name but a few. Many of these approaches are based on similar sampling algorithms and in this work we have elected to apply the standard SMOTE algorithm as it has been effectively applied in many domains.

## III. THE JAZZ DATASET

IBM Jazz is a fully integrated software development tool that automatically captures software development processes and artifacts. The Jazz repository contains real-time evidence that allows researchers to gain insights into team collaboration and development activities within software engineering projects [1, 7]. The Jazz repository artifacts include work items, build items, change sets, source code files, authors and comments. A work item is a description of a unit of work, which is categorized as a task, enhancement or defect. A build item is compiled software to form a working unit. A change set is a collection of code changes in a number of files. In Jazz a change set is created by one author only and relates to one work item. A single work item may contain many change sets. Source code files are included in change sets and over time can be related to multiple change sets.

One of the challenges associated with working with the Jazz repository is that the data contains holes and misleading elements which cannot be removed or identified easily. This is because the Jazz environment has been used within the development of itself; therefore many features provided by Jazz were not implemented at early stages of the project. This sparseness of the data has driven the decision to focus on using software metrics as the predictor attributes. Whilst features of the Jazz environment may not have been present during early phases of development, there has always been source code and therefore a consistent set of data can be created.

## IV. THEORETICAL FOUNDATIONS

Software metrics have been generated in order to deal with the sparseness of the data. Metric values can be derived from extracting development code from software repositories. Such metrics are commonly used within model-based project management methods. Software metrics are used to measure the complexity, quality and effort of a software development project [8-12]. In the Jazz repository each software build contains change sets that indicate the actual source code files that are modified during the implementation of the build. Source code metrics for each file are computed using the IBM Software Analyzer tool. The builds after state was utilized in order to ensure that the source code snapshot represented the actual software artifact that either failed or succeeded.

The Jazz repository consists of various types of software builds. Included in this study were continuous builds (regular user builds), nightly builds (incorporating changes from the local site) and integration builds (integrating components from remote sites). As a result the following basic, average basic, dependency, complexity, cohesion and Halstead software metrics were derived from the source code files for each build:

- Basic Software Metrics:
  - Number of Types Per Package, Number of Comments, Lines of Code, Comment/Code Ratio, Number of Import Statements, Number of Interfaces, Number of Methods, Number of Parameters, Number of Lines, Average Number of Attributes Per Class, Average Number of Constructors Per Class, Average Number of Comments, Average Lines of Code Per Method, Average Number of Methods, Average Number of Parameters.
- Dependency Metrics:
  - Abstractness, Afferent Coupling, Efferent Coupling, Maintainability index, Instability, Normalized Distance.
- Complexity Metrics:
  - Average Block Depth, Average Cyclomatic Complexity.
- Cohesion Metrics:
  - Lack of Cohesion 1 (LCOM1), Lack of Cohesion 2 (LCOM2), Lack of Cohesion 3 (LCOM3).
- Halstead Metrics:
  - Number of Operands, Number of Operators, Number of Unique Operands, Number of Unique Operators, Program Volume, Difficulty Level, Effort to Implement, Number of Delivered Bugs, Time to Implement, Program Length, Program Level, Program Vocabulary Size.

### A. Synthetic Minority Over-sampling TEchnique (SMOTE)

When working with real world data it is often found that data sets are heavily comprised of "normal" instances with only a small percentage representing interesting findings. As a result the "abnormal" instances have a negative impact on a models' performance as they have a greater probability of misclassification using data mining methods [2, 13]. Data instances that introduce noise within the data are often found within the minority class [14, 15]. In order to overcome this limitation synthetically under-sampling the majority class may improve a classifiers' performance. However, in doing so valuable data may be lost and model over-fitting may occur, resulting in majority instances being wrongly classified as minority instances when new, unseen data is presented to the classifier model that was induced [14]. Another solution is to provide the classifier with more complete regions within the feature space via creation of new instances that are synthesized form existing data instances.

SMOTE enables a data miner to over sample the minority class to achieve potentially better classifier performance without loss of data [2, 13]. While other over-sampling methods exist, such as Rippers Loss Ratio and Naive Bayes methods, SMOTE provides better levels of performance as it generates more minority class samples for a classifier to learn from, thereby allowing broader decision regions and coverage [13]. SMOTE has been utilized within the software research community and compared with other sampling techniques in software quality modeling (random under-sampling, random oversampling, cluster-based oversampling and Borderline-SMOTE) and has yielded encouraging results [5, 8]. SMOTE has also been applied as a sampling strategy for software defect prediction where data sets from NASA software project data sets [10,16-18] and fault-prone module detection using the MIS telecommunication systems [24]. For this work SMOTE is applied as a supervised instance filter using the Weka [19] machine learning workbench.

In order to avoid the over-fitting problem while expanding minority class regions SMOTE generates new instances by operating within the existing feature space. New instance values are derived from interpolation rather than extrapolation, so they still carry relevance to the underlying data set. For each minority class instance SMOTE interpolates values using a k-nearest neighbor technique and creates attribute values for new data instances [8]. For each minority data a new synthetic data instance (I) is generated by taking the difference between the feature vector of I and its nearest neighbor (J) belonging to the same class, multiplying it by a random number between 0 and 1 and then adding it to I. This creates a random line segment between every pair of existing features from instances I and J, resulting in the creation of a new instance within the data set [13]. This process is repeated for the other k-1 neighbors of the minority instance I. As a result SMOTE generates more general regions from the minority class and decision tree classifiers are able to use the data set for better generalizations.

### B. Hoeffding Tree

The Hoeffding tree is an incremental decision tree induction method. Using the Hoeffding bound, it ascertains the number of instances that are needed to split a given (decision) node of a tree and operates within a certain precision that can be predetermined [3]. This method has potential in terms of predicting future outcomes of software builds with high accuracy while working with real-world data. Rather than using training and test sets, instances are represented as streams. The Hoeffding tree is commonly used for classifying high speed data streams. The algorithm that it uses generates a decision tree from data incrementally by inspecting each instance within a stream without the need to store instances for later retrieval. The tree resides in memory during each iteration and stores information in its branches and leaves, potentially growing from "learning" every new instance. The decision tree itself can be inspected at any time during the streaming process. The quality of the tree itself is comparable to that used by traditional mining techniques, even though instances are introduced in an incremental manner.

Just as with traditional decision tree learners, the Hoeffding tree is easy to interpret, making it easier to understand how the model works. In addition to this, decision tree learners have proven to provide accurate solutions to a wide range of problems that are based on multi-dimensional data. For Hoeffding trees each node of a decision tree undergoes a test which may result in it being split into two or more child nodes and sending each instance down a relevant branch to its destination child node, depending on the values of its attributes. The split test is implemented through the use of the Hoeffding bound which is expressed as:

$$\epsilon = \sqrt{\frac{R^2 \ln\left(\frac{1}{\delta}\right)}{2n}} \qquad (1)$$

The Hoeffding bound expressed in (1) above states that with confidence (1- $\delta$), the population mean of R lies in the interval, [$\bar{R}$ -$\epsilon$, $\bar{R}$ +$\epsilon$], where $\bar{R}$ is the sample (observable) mean of the random variable R. In the context of decision tree induction R refers to information gain. The Information gain function ranges in value from 0 to $log_2 c$, where c is the number of classes. Since c=2 in the mining problem that we undertake (since only the outcomes, *success* and *failure* are possible), R reduces to 1. The variable n refers to the number of data instances seen up to the point that the test was carried out. The bound holds is true irrespective of the underlying data distribution generating the values and only depends on a range of values, number of observations made and a split confidence level. The Hoeffding tree uses the Hoeffding bound to determine whether an existing (leaf) node should be split as follows. Suppose that after n data instances have arrived, the difference in information gain between the two highest ranking attributes $X_a$ and $X_b$ with $\Delta \bar{G} = \bar{G}(X_a) - \bar{G}(X_b) > \tau$ (i.e. $X_a$ is the attribute with the highest information gain), then with confidence (1- $\delta$), the Hoeffding bound guarantees that the correct choice to spilt the given leaf node is attribute $X_a$ if $\Delta \bar{G} > \epsilon$, where $\tau$ is a tie threshold parameter.

In this research we use the Hoeffding tree implementation from MOA [13], a real time analytics tool for data streams was used for mining data streams.

## V. EXPERIMENTAL STUDY

The original software metric data set consists of 199 Jazz build instances. From these instances there are 127 successful builds and 72 failed builds. Build instances are sorted by date to ensure accurate simulation of a development team working over time. SMOTE is then applied twice at 900%, increasing the number of instances to 1,990 (1270 successful builds and 720 failed builds). The first application increases the number of minority class instances (failed builds) and the second application increases the temporarily "new" number of minority class instances (successful builds). The instances are then encoded into data streams which are utilized by the Hoeffding tree for the data mining process. Three parameters were set for the tree induction. The Hoeffding tree uses a grace period parameter which stipulates the frequency with which checks for leaf node splits are carried out, the greater the value the higher the efficiency of the process. We use a setting of 200 for the grace period parameter. The tie threshold parameter, $\tau$ that controls the degree of splitting, was set to 0.05.

Presented in Figure 1 is the classification accuracy obtained with the use of after state metrics for builds. The classification accuracy at the start of the time series was 65.2% and at the end of the stream the accuracy grew to 80.25%. The average overall accuracy over the entire time series was 70%. This indicates that the there is potential for the accuracy of prediction to improve as more real data emerges.

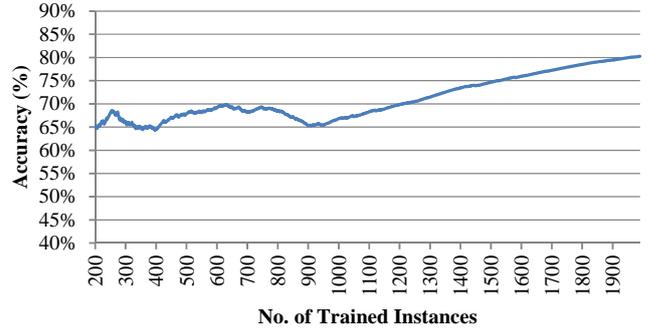

Figure 1. Hoeffding Tree Overall Classification Accuracy.

The initial instability in classification accuracy is an interesting phenomenon, given the initial grace period of 200 builds is intended to provide stability in the emerging model. Upon examination of the synthetic data it can be observed that the data maintains comparable instances of each class up until 900 builds. After 900 builds, the data contains an increasing proportion of successful builds. This is shown in Figure 2.

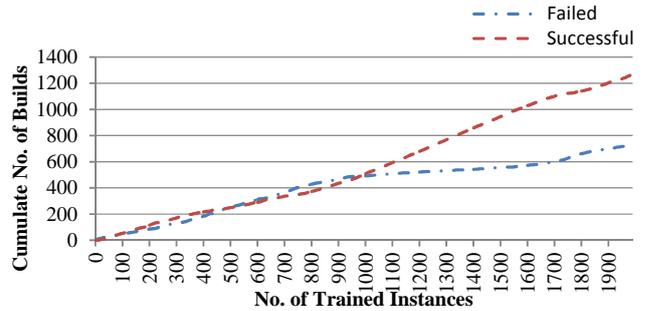

Figure 2. Build Distribution over Time

Figure 3 presents the classification accuracies of successful builds. It is observed that the general trend for classifying success initially declines to reach a minimum at approximately 900 instances, after which there is a gradual improvement that appears to be trending towards a stable value of around 80%. Figure 4 displays the sensitivity ratings for successful builds over time. For successful builds the accuracy at the beginning of the data stream time series was 66.38% and ended with 79.1% (with an average of 64%).

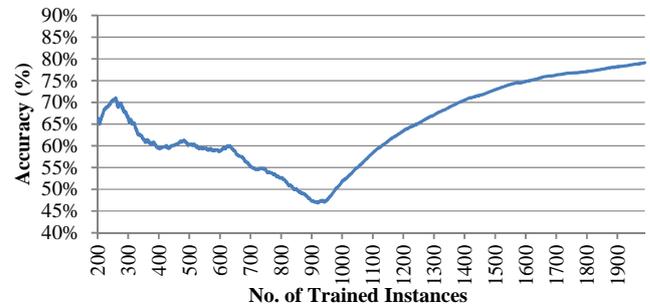

Figure 3. Hoeffding Tree Classification Accuracy for Successful Builds.

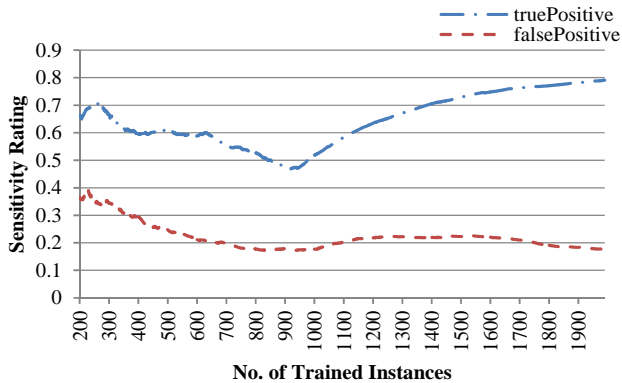

Figure 4. Hoeffding Tree Sensitivity Measurements for Successful Builds.

Presented in Figure 5 are the classification accuracies over time for failed builds and the corresponding sensitivity ratings for failed builds are presented in Figure 6. Classification accuracy for failed builds started at 63.5% and at the end of the time series was 82.2% (with an average of 78%). The false positive values between 700 to 1000 trained instances appear to peak when classifying failed builds, due to over-fitting the model at earlier time segments. The false positive value then proceeds to decrease over time

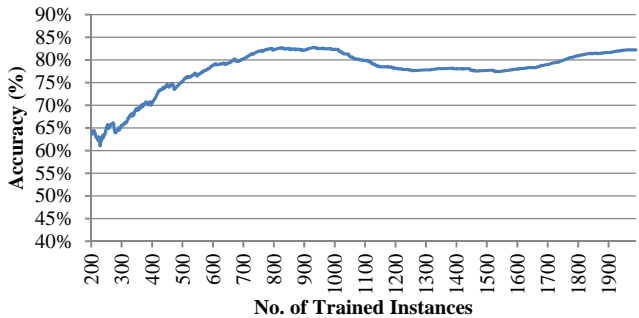

Figure 5. Hoeffding Tree Classification Accuracy for Failed Builds.

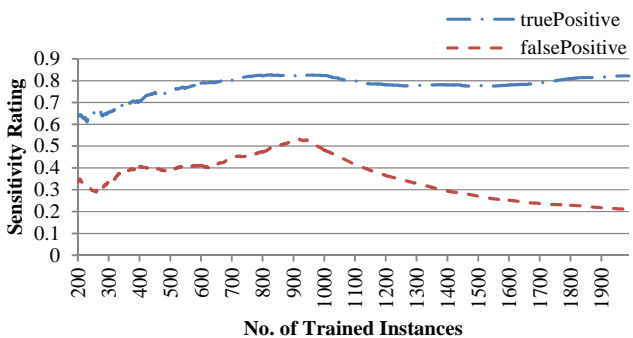

Figure 6. Hoeffding Tree Sensitivity Measurements for Failed Builds.

Interestingly, the distribution of build instances across the two classes has marginal impact on the overall classification accuracy when compared to the impact it has on the individual classes themselves. When the distribution of classes in the synthetic data is roughly equal there is an increase in the classification accuracy of failed builds that is accompanied by a decrease in classification accuracy of successful builds. This seems at odds with the observation of previous work [21, 22, 30] that suggests that failed builds are harder to classify than successful builds. This work suggests that failed builds may be harder to classify when there is a significantly larger number of successful builds that dominate the classification model.

Figure 7 illustrates the final decision tree using the Hoeffding Tree stream mining technique on the extended RSA after state software metrics data set. In this case the tree is larger than the previous software metric based Hoeffding tree, with a depth of 7. Upon inspecting the tree there are common sense classifications being made, for example a higher number of interfaces tend to be associated with failure. This is intuitive because if there are too many Java interfaces it can become tedious when debugging an error as the actual implementation of the error may be in an obscure location. Interfaces also add to the collection of files within the system and if an interface is "dead" (not used) and not removed it leads to a less elegant system design. The number of interfaces has a direct influence on dependency metrics, i.e. Abstractness.

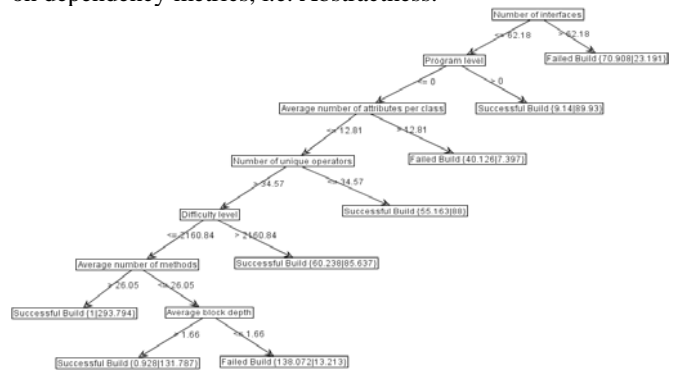

Figure 7. Final Hoeffding Tree for After State Software Metrics.

## VI. LIMITATIONS AND FUTURE WORK

The Jazz repository contains holes and misleading elements which cannot be removed or identified easily. There is a great challenge in dealing with such inconsistency and the methodology has adopted an approach that delves further down the artifact chain than most previous work using Jazz. It is a premise that the early software releases were functional, so whilst the project "meta-data" may be missing details (such as developer comments) the source code should represent a stable system that can be analyzed to gain insight regarding the development project. Even when comparing to other Jazz studies there are concerns over validity that arise from the possibility of different extraction techniques being applied. However, the approach for creating a predictive model by mining data streams that relate to software data can be applied to other repositories and as such is a generalizable process. Similarly, the process of using the predictive model to identify build outcome risk and proactively manage the build scope and activities is equally applicable to other projects. The actual prediction models are likely to be very different for other projects, but the techniques for developing them are entirely generic. Other limitations from this study are products of the relatively small sample size of build data from the Jazz project combined with the sparseness of the data itself. For example, the ratio of metrics (42) to builds (199) is such that it is difficult to truly identify significant metrics. Even though a sampling technique (SMOTE) is applied to increase the number of instances, it is not possible to assess the extent to which the generated data reflects real-world data as there is the likelihood of unpredictable events in software development projects.

VII. CONCLUSIONS

The goal of synthetically generating data was to explore what might happen if there was more data available for mining, more specifically to see if classification of builds improved with more data. While the use of SMOTE may not be a "true" representation of future real world data, it does however interpolate values between existing instances to generate new data that may be considered at representative of existing data. This provides insights into what may occur if there were no "new" anomalies encountered during the project. This may not be entirely realistic given that the causes of failure are not predictable and that new failure modes are likely to appear over time. From previous data mining experiments it was observed that build failure metrics were often overlapped in value with those of successful builds, thus challenging the ability of a classifier to distinguish between these two types of build outcomes. This indicates that if more data is available accuracy for classifying builds may improve over time. The results obtained during this phase support other studies where software build outcome prediction accuracy and stability both increased when adopting the use of SMOTE [10, 11, 20] on other project software metrics. While this research has gone some way to addressing the challenges associated with data mining software repositories, there is still much potential for future work in understanding evolving success and failure patterns found within the SDLC.